\documentclass[doublecol]{epl2} 
\usepackage{subfig}
\usepackage{amsmath}
\usepackage[normalem]{ulem}

\title{Entropy production for velocity-dependent macroscopic forces: the problem of
  dissipation without fluctuations.} \shorttitle{Entropy production for macroscopic velocity-dependent forces} 

\author{L. Cerino\inst{1} \and A. Puglisi \inst{1}}
\shortauthor{L. Cerino \etal}
	
\institute{                    
  \inst{1} Istituto dei Sistemi Complessi - CNR and Dipartimento di Fisica, Universit\`a di Roma Sapienza, P.le Aldo Moro 2, 00185, Rome, Italy
}
\pacs{05.70.Ln}{Thermodynamics, nonequilibrium}
\pacs{05.10.Gg}{Stochastic models in statistical physics and nonlinear dynamics}

\abstract{In macroscopic systems, velocity-dependent phenomenological
  forces $F(v)$ are used to model friction, feedback devices or
  self-propulsion. Such forces usually include a dissipative component
  which conceals the fast energy exchanges with a thermostat at the
  environment temperature $T$, ruled by a microscopic Hamiltonian
  $H$. The mapping $(H,T) \to F(v)$ - even if effective for many
  purposes - may lead to applications of stochastic thermodynamics
  where an {\em incomplete} fluctuating entropy production (FEP) is
  derived. An enlighting example is offered by recent macroscopic
  experiments where dissipation is dominated by solid-on-solid
  friction, typically modelled through a deterministic Coulomb force
  $F(v)$. Through an adaptation of the microscopic Prandtl-Tomlinson
   model for friction, we show how the FEP is dominated by the heat released
  to the $T$-thermostat, ignored by the macroscopic Coulomb
  model. This problem, which haunts several studies in the literature,
  cannot be cured by weighing the time-reversed trajectories with a
  different auxiliary dynamics: it is only solved by a more accurate
  stochastic modelling of the thermostat underlying dissipation.}

\begin{document}

\maketitle

\section{Introduction}

Since its infancy, non-equilibrium statistical mechanics has been based on models with coarse-grained forces~\cite{mazenko2006,Castiglione2008}. 
Indeed the origin of thermostats and
non-conservative forces is a reduction of the description from a
larger Hamiltonian system, where many degrees of freedom have been
projected out~\cite{zwanzig}.

Dissipation is an essential ingredient in non-equilibrium
systems~\cite{K66}, consisting in a net - or average - transfer of
energy from the degrees of freedom we are observing (i.e. the system),
to a large and often hidden reservoir - the environment - which has a
lower energy density.  For large macroscopic systems the
transfers of energy in the opposite
direction are very unlikely~\cite{Lebowitz1994}. Mesoscopic systems, where such
fluctuations are non-negligible, are the subject of stochastic
thermodynamics (ST) whose study has received a great impulse in the
last 20 years~\cite{Seifert2005,Sekimoto2010}. A central issue in ST
is relating fluctuations of currents, such as energy flows, to
the fluctuating entropy production (FEP)~\cite{Lebowitz1999}.

A major contribution of the present Letter is to provide a neat
example where such dissipation-FEP connection is dramatically broken
when the macroscopic model ignores the
microscopic fluctuations. We take into detailed consideration the force
acting between two sliding solid bodies, that is the so-called dry
friction, which in macroscopic systems (e.g. on scales larger than few
millimeters) is well described by the law of Coulomb
friction~\cite{GPT13,manacorda,VMUZT13}. 
In our analysis it becomes clear that the Coulomb  model is
too much coarse-grained and, for this reason, it neglects the dominant
contribution to the FEP~\cite{pigolo,Crisanti2012}.  The proper
thermodynamics is restored by taking into account the underlying
thermostat.

Apparently, this problem has been overlooked in the recent
literature. For instance, entropy production of Coulomb friction is usually neglected
in Langevin models~\cite{BTC11}. In~\cite{MR12,MR14} a Langevin
equation with feedback is considered, generalizing a model for atomic
manipulation of macromolecules~\cite{KQ04}. When velocity-dependent,
the feedback mechanism in these models dissipates energy:
in~\cite{KQ04} it is explicitly recognised that it acts as a {\em
  refrigerator}. However the application of stochastic thermodynamics,
even with the help of artificial ``conjugate'' time-reversed dynamics,
leads to a formula for FEP which does not take into account the heat flowing to the
low-temperature thermostat. Similar problems affect a macroscopic
velocity-dependent force~\cite{GC13} which accounts for both friction
and self-propulsion in an active matter model~\cite{active2}.  In all
these cases the error does not reside in the recipe of stochastic
thermodynamics, but rather in the incomplete modelling of
dissipation.




\section{Levels of coarse-graining and dissipation}

In the present paper we consider two levels of coarse-graining: C1 and
C2. The first level, C1, is the widely adopted reduction of a large
Hamiltonian system into a sub-system (the part of interest) plus a
thermostat that obeys a  much simpler
dynamics~\cite{zwanzig}. The second coarser level C2 is
useful to describe macroscopic systems, when the
perturbation acts on scales much larger than the
microscopic ones.

\begin{figure}
\centering
	\includegraphics[width=0.4\columnwidth,clip=true]{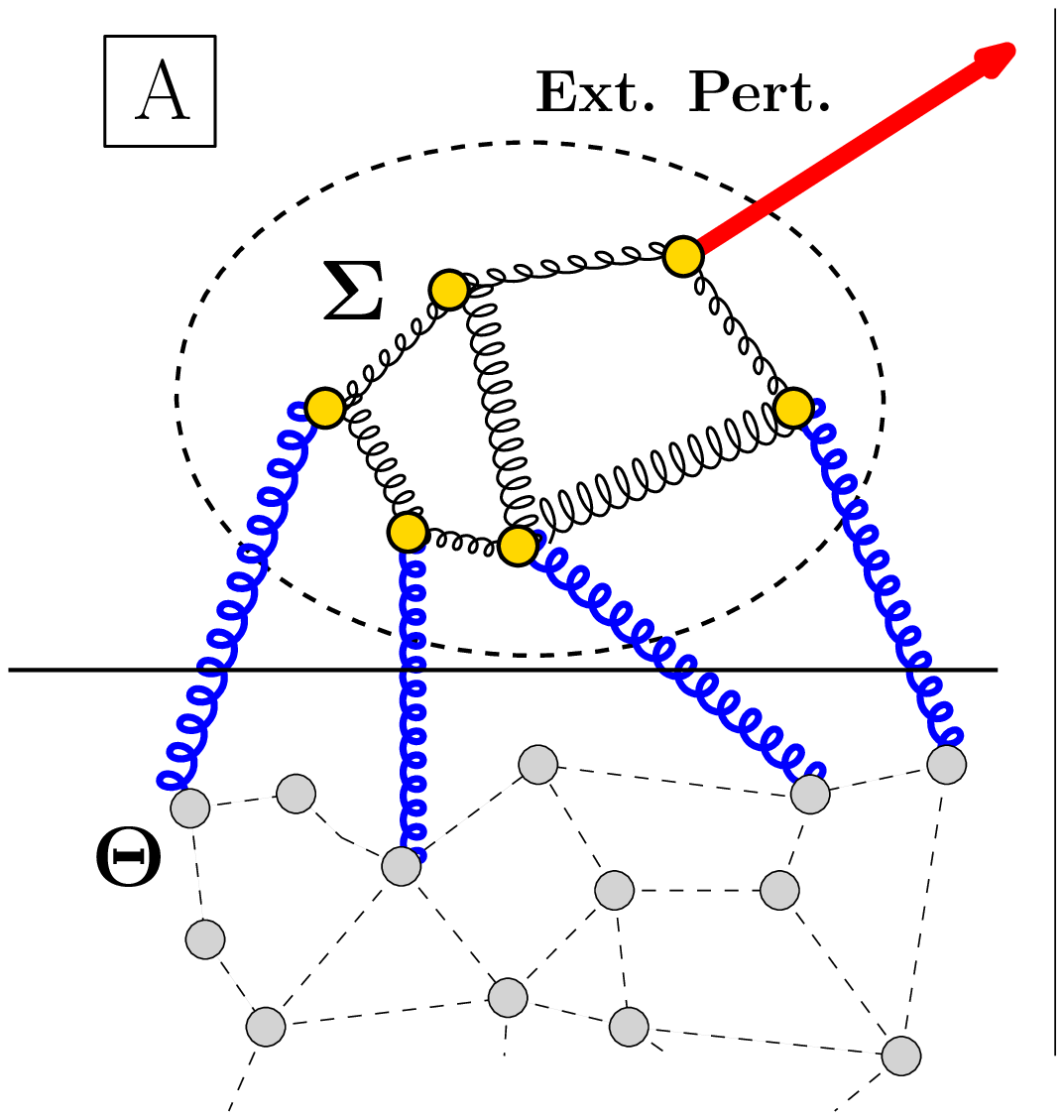}
\includegraphics[width=0.4\columnwidth, clip=true]{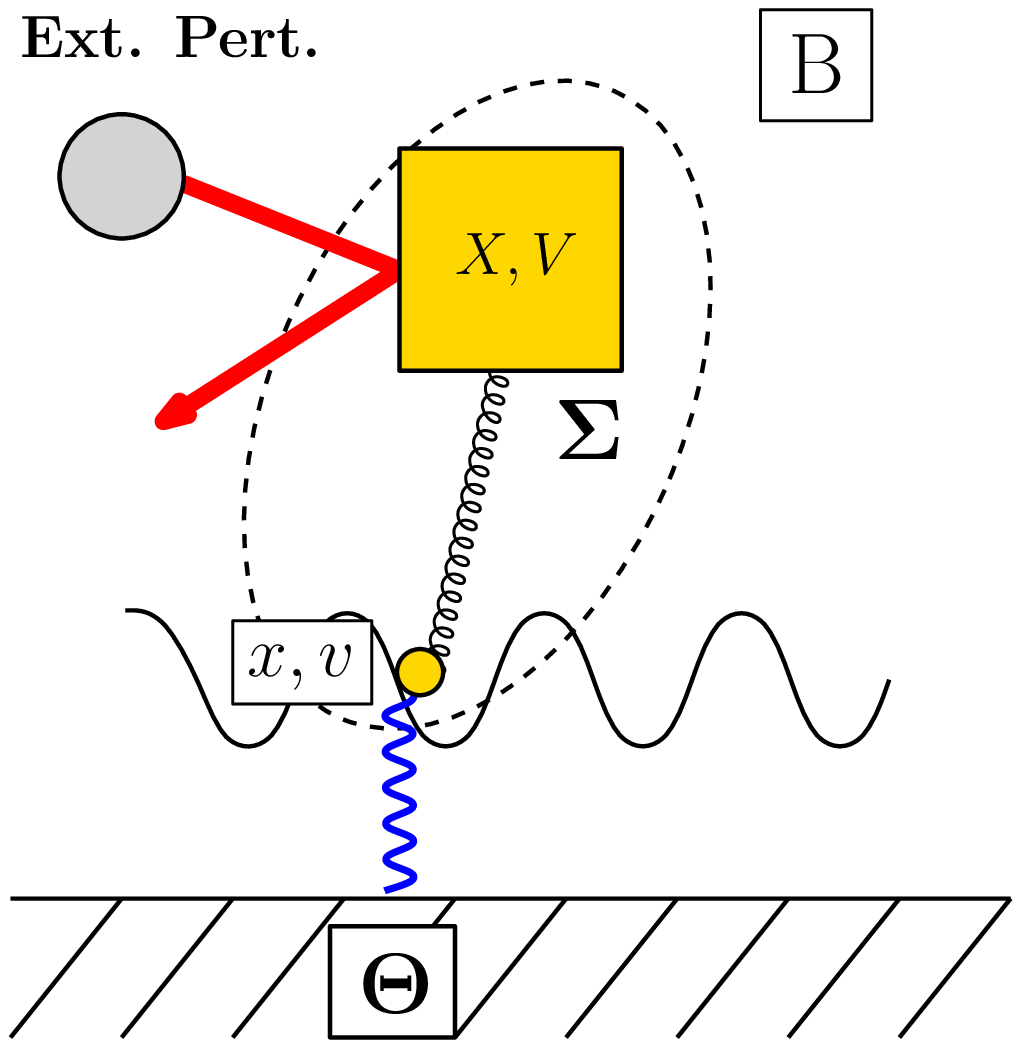}
\caption{A: sketch of the generic model we discuss in the section
  ``Levels of coarse-graining and dissipation'', Eqs.~\eqref{eq:master0} and~\eqref{eq:master}. B:
  sketch of the PT model discussed in the section ``The
  Prandtl-Tomlinson Model'', Eq.~\eqref{eq: 1}. }
\label{fig:sk}
\end{figure}

Figure~\ref{fig:sk}A depicts a general scheme to discuss C1. A
Hamiltonian system is composed of two interacting sub-systems
$\Sigma$ and $\Theta$: a point in the complete phase (``zero level'') space is $\Gamma_0
=(\Gamma_\Sigma,\Gamma_\Theta)$.  In the limit where $\Theta$ is much
larger than $\Sigma$, one can consider $\Theta$ as a ``thermostat'', since
its energy is barely affected by its interaction with $\Sigma$. If
need be, an external perturbation is applied to some of the degrees of
freedom of $\Sigma$: the nature and formal details of the perturbation
are discussed later. 
We consider a probabilistic description where the phase space
position $\Gamma_0$ is distributed with some probability density
$P_0(\Gamma_0,t)$ at time $t$. This density obeys an equation of the
kind

\begin{equation}
\label{eq:master0}
\frac{\partial P_0(\Gamma_0,t)}{\partial t} = [L_0(\Gamma_0)+L_{ext}(\Gamma_\Sigma,t)] P_0(\Gamma_0,t),
\end{equation}

where $L_0$ is the Liouville operator associated to the
total Hamiltonian $H_0(\Gamma_0)$ and $L_{ext}$ represents the external
perturbation, which can be deterministic, stochastic, time-dependent
or not, etc.
The first level of coarse-graining, C1, consists in focusing on
$\Gamma_1 \equiv \Gamma_\Sigma$ alone, by replacing Eq.~\eqref{eq:master0} with the
following~\cite{zwanzig}:

\begin{subequations} \label{eq:master}
\begin{align}
\frac{\partial P_1(\Gamma_1,t)}{\partial t} &= [L_1(\Gamma_1)+L_{ext}(\Gamma_1,t)] P_1(\Gamma_1,t),\\
L_1(\Gamma_1) &=L_H(\Gamma_1) + L_T(\Gamma_1).
\end{align}
\end{subequations}

In Eq.~\eqref{eq:master}, the $L_H$ operator is the Liouville operator
associated to the Hamiltonian $H_{}(\Gamma_1)$ of the system $\Sigma$
alone, i.e. the internal dynamics of the system of interest and $L_T$
is the operator describing - in some simplified form - its coupling to
$\Theta$. A common and convenient choice for $L_T$ is a stochastic
operator: for instance, if memory effects can be neglected, $L_T$
takes the form of a Markovian master equation operator. Its transition
rates, in order to reflect the invariance of $L_0$ under
time-reversal, must satisfy detailed balance with respect to the Gibbs
measure defined by Hamiltonian $H_{}$ and temperature $T$.

A tool which has been widely used for a formal characterization of
time-reversal invariance (or variance), at the level of single
trajectories, is the so-called action functional~\cite{Lebowitz1999}:
\begin{equation} \label{eq:ls}
W_i(t)=\ln \frac{p(\{\Gamma_i(s)\}_0^t|\Gamma_i(0))}{p(\{\epsilon \Gamma_i(t-s)\}_0^t|\epsilon\Gamma_i(t))},
\end{equation}
where $i$ can be $0$ or $1$ depending on the level of description. The
numerator in~\eqref{eq:ls} represents the probability, conditioned to
the state at time $0$, of a trajectory $\Gamma_i(s)$ with
$s\in(0,t)$. The denominator is the probability, conditioned to the
time-reversed final state, of the time-reversed trajectory:
$\epsilon$ is a diagonal operator which leaves unchanged all positions
and changes sign to all velocities. It is meant, of course, that the
conditional probability for the trajectory at level $i=0$ ($i=1$) is
generated by Eq.~\eqref{eq:master0} (Eq.~\eqref{eq:master}). When
$L_{ext}=0$ one has $W_0(t) \equiv 0$ and $W_1(t) =
\frac{H_{}(\Gamma_1(0))-H_{}(\Gamma_1(t))}{k_B T} $
(detailed balance). The cases where both $L_{ext}$ and $L_T$ are
deterministic are usually discussed in the context of phase-space
contraction~\cite{GC95}. Several physical examples have been given in
the literature\footnote{The reader is warned that we
do not intend to exhaust in a few sentences the huge field of
FEP, our aim is to summarize a few basic
observations. Those are made rigorous under more precise hypotheses, and,
possibly, with the addition of the so-called ``boundary terms''~\cite{bordi,bordi2}} where the action functional takes the form
$\mathcal{W}_{ext}(t)/{k_B T}$,
with $\mathcal{W}_{ext}(t)$ the work done by the non-conservative
forces during the trajectory. Another situation commonly discussed
is when $L_{ext}$ represents the interactions with a second bath at
a temperature $T' \neq T$: in that case the action functional takes
the form
$\mathcal{Q}(t)\left|\frac{1}{T}-\frac{1}{T'}\right|$,
with $\mathcal{Q}(t)$ the energy transferred from the external
thermostats into the system during the trajectory. Both examples
suggest a strong analogy between the action functional and
thermodynamic entropy production, where the energy flowing to the
thermostat, i.e. the dissipation, enters explicitly. In stochastic
thermodynamics the formulation we choose in
Eq.~\eqref{eq:ls} is referred to as FEP of the
medium~\cite{Seifert2005}.

When the external perturbation $L_{ext}$ acts on space-time scales
much larger than those dictated by $L_H$ and $L_T$, it is convenient
and common to scale up the description to a ``macroscopic''
level, what we call the C2 coarse-graining. This operation is usually
achieved by phenomenological considerations, cases where it can be
rigorously carried on being rare. Only those degrees of freedom which
are relevant at large scales, $\Gamma_{2}$, are retained and the
evolution of their probability takes the form
\begin{equation} \label{eq:master2}
\frac{\partial P_2(\Gamma_2,t)}{\partial t} = [L_2(\Gamma_2) + L_{ext}(\Gamma_2,t)] P_2(\Gamma_2,t).
\end{equation}
The $L_2$ operator represents the contraction of microscopic
Hamiltonian ($L_H$) and thermostat ($L_T)$ parts, and it is often
non-conservative and deterministic. Indeed, the aim of C2 is to get
a fair description of the trajectories at a macroscopic resolution
where energy is dissipated and fluctuations are usually negligible. The result is that
$L_2$ fairly accounts for dissipation, but may violate detailed balance, a required symmetry in the absence of $L_{ext}$.  In addition, in view of the huge difference of energy scales, the probability that the time-reversal of a typical trajectory is observed is exceedingly small: for this
reason, in many common cases~\cite{KQ04,BTC11,GPT13,GC13,MR14}, $L_2$
does not describe properly those trajectories and strongly twists $W(t)$.
  Generalisations $\tilde{W}(t)$ of the action functional
$W(t)$ have been proposed (see for
instance~\cite{jarzy2006,ford2012,MR14}) where the probability
weighing the reversed trajectory, i.e. that appearing at the
denominator of Eq.~\eqref{eq:ls}, is replaced by a different
probability, generated by an {\em auxiliary dynamics}. Unfortunately
this ad hoc prescription changes the physical meaning (and the
accessibility in experiments) of the action functional and usually does
not solve the discrepancy. The only way to get an action functional
which properly represents the thermodynamics of the system is to
include, in $L_2$, the fluctuations which are conjugate to the
modelled dissipation restoring the condition of detailed balance in absence of $L_{ext}$~\cite{hanggi09}.  In the next Sections we discuss in details a
clear example to understand this scenario.

\section{The case of Coulomb friction}

A particularly interesting  macroscopic dissipative force (i.e. of the kind of $L_2$) is the so-called dry or
solid-on-solid friction: it describes the force against
relative sliding between two solid surfaces at contact~\cite{persson}. In its simplest
and oldest form, which is considered a fair approximation at the C2 level of coarse-graining,
it reads
\begin{equation} \label{eq:coulomb}
F_C(V) = -\Delta \sigma(V),
\end{equation}
where $\Delta=\mu F_N$ is a positive force proportional, through the
(dynamical) friction coefficient $\mu$ to the normal force $F_N$, and
$\sigma(v)$ is $+1$,$-1$,$0$ if $v>0$, $v<0$ or $v=0$ respectively.
When the body is at rest also the static friction force should be 
considered: however its role is marginal in this context because the 
external driving typically gives strong impulsive forces to the 
sliding mass.

In the last decades the study of solid friction, taking also advantage
of experimental techniques at the micro and nano scales, has refined
dramatically the simple law in
Eq.~\eqref{eq:coulomb}~\cite{VMUZT13}. It is known that it should be
modified to take into account thermal effects, aging of contacts,
dependence upon $V$, and much more. Notwithstanding those progresses,
Eq.~\eqref{eq:coulomb} remains useful in simple macroscopic
situations. An example where it fairly describes experimental results
is in~\cite{gnoli,GSPP13,GPT13}: a solid macroscopic rotator is in
contact with a fluidized granular gas made of spherical beads of mass
$M_g$ and granular temperature $T_g$ (see Eq.~\eqref{eq:rates}
below for an operative definition). The beads hit the solid body and
excite its rotation, which is then damped by solid friction in the
ball bearings allowing rotation. In the following we also use the name
``tracer'' to indicate the rotator, and we use $X$ and $V$ to mean its
(angular) position and velocity respectively. When the granular gas is
dilute the collision are described by a non-continuous Markov process
with transition rates dictated by the collisional
kinetics~\cite{talbot2,KSSH15}.  The master-equation for $P(X,V,t)$ then is
equivalent to
Eq.~\eqref{eq:master2} with $\Gamma_2 \equiv (X,V)$ and
\begin{subequations} \label{eq:langevin}
\begin{align} 
L_2 \cdot &=-\partial_{X} [V \cdot] -\partial_{V} \left[\frac{F_C(V)}{M} \cdot\right],\\
L_{ext} P(X,V,t)  &= \int dU P(X,U,t) k(U \to V) \label{eq:grmarkov}\\
  - \int dU &P(X,V,t)  k (V \to U). \nonumber
\end{align}
\end{subequations}
A simplified form of the
transition rates, used in some theoretical studies~\cite{lorentz} and convenient for its
simplicity, is the following
\begin{equation} \label{eq:rates}
k(V\to V')=\tau^{-1}\left(\frac{M_g+M}{2 \sqrt{M_g}}\right)\frac{e^{-M_g u^2(V,V')/2(k_B T_g)}}{\sqrt{2\pi k_B T_g}},
\end{equation}
where $T_g$ is the ``temperature'' of the granular gas and
\begin{equation}
u(V,V')=\frac{M_g+M}{2 M_g}(V'-V) +V.
\end{equation}
In the above transition rates we have not considered the inelasticity
of collisions, which is indeed negligible with respect to the
dissipation due to $F_C(V)$ when collisions are not too frequent
(``rare collision limit'', discussed in~\cite{talbot2}).
To give an idea of the
energy, space and time scales, it should be considered that in the experiments the
mass of the rotator is $\sim 5$ g, the diameter of the beads
is $\sim4$ mm and their mass is $\sim 10^{-1}$ g, while their average speed is
$\sim 10^2$ mm/s, that is an average kinetic energy of the order of
$\sim 10^{-6}$ J, leading to a $T_g$ of the order
of $\sim 10^{17}$ K.
Other experiments have been performed where Coulomb force $F_C(V)$ is
coupled to time-dependent external perturbations
~\cite{mahadevan,daniel,eglin}. Those experiments have also triggered
the interest of many theoreticians who studied the problem with
different kinds of noise~\cite{buguin,fleishman,baule}. 

One of the less studied aspects of Eq.~\eqref{eq:langevin} is its
behavior under time-reversal. There is an evident obstacle in doing
that: if a trajectory $\{V(s)\}_0^t$ between times $0$ and $t$
solves Eq.~\eqref{eq:langevin} with a given noise realization, there
is no way (by means of any other noise realization) that the
time-reversed trajectory $\{-V(t-s)\}_0^t$ satisfies the same
equation. Indeed all the parts of the dynamics where only friction is
acting (decreasing $|V|$) are mapped, by time-reversal, to
trajectories taking energy (increasing $|V|$), which are forbidden
by $F_C(V)$. A consequence of this observation is that the action
functional Eq. (\ref{eq:ls}) cannot be properly defined.
Nonetheless, we can always empirically define the FEP of a system coupled to several reservoirs as the sum of the
energies that the system exchanges with each thermostat, divided by
the temperature of the thermostat. Since our system seems
coupled to a single thermostat (i.e. the grains) this definition
leads to
\begin{equation}\label{eq:current}
\tilde{W}(t)=-\sum_{i=1}^{N_c}\frac{\delta E_i}{k_B T_g},
\end{equation}
where $N_c$ is the total number of collisions occurring in the
interval $[0,t]$ and $\delta E_i = \frac{M}{2}(V_i'^2-V_i^2)$ the
energy gained in the $i$-th collision.  Another possibility is to
choose $\tilde{W}(t)$ by modifying the action functional~\eqref{eq:ls}: the probability of the
time-reversed trajectories is generated with an auxiliary dynamics
obtained by inverting the sign in front of the Coulomb force~\cite{jarzy2006,ford2012,MR14}. With
this prescription the generalized action functional takes the exact
form of Eq. (\ref{eq:current}) (see the analogous calculation in the next Section).

When $M_g\ll M$, it is possible~\cite{vK61} to approximate
Eq. (\ref{eq:grmarkov}) with
$L_{ext}~\cdot=M^{-1}\partial_{V}[\gamma_g V~\cdot] +
M^{-2}\gamma_g T_g \partial^2_{V}[~\cdot~]$: in such a limit
\cite{dGen05,H05} the paradox vanishes, because noise acts
continuously and some noise realization that sustains the reversed
trajectory can always be found (with different probability, of
course). On the other hand, the limit is singular in the sense that
Eq.~\eqref{eq:langevin} with white noise is a potential equation and
satisfies detailed balance: the stationary state is an equilibrium
state in which the action functional has zero average and the physical meaning
of dissipation is totally lost~\cite{BTC11}. Nevertheless, the
generalized action functional obtained with the above ``corrected''
sign prescription takes the form, up to boundary terms,
$\tilde{W}(t)=-\Delta \int_0^t dt' |V(t')|/T_g$,
i.e. minus the work done by the friction force divided by the
temperature of the thermostat. Since in the stationary state this work
is on average equal to the energy exchanged with the thermostat, this
quantity is in agreement with our intuitive definition of entropy and
with Eq.~\eqref{eq:current}. Despite the apparent coherence of the
above proposed solutions we will show, in the next Section, that the
results are incomplete, mainly because the coarse graining is hiding the thermostat responsible for the largest part of the FEP.
It should be noticed that measures of the fluctuations of these and
other physical currents (e.g. angle spanned by the rotator in a time
$t$) have been performed, finding interesting large deviations
properties~\cite{sarra}.  In order to understand the connection between measured
macroscopic currents and the appropriate FEP, in the
next Section we will resort to a more fundamental model (C1 level).

\section{The Prandtl-Tomlinson Model}
A good
candidate reveals to be the so-called Prandtl-Tomlinson (PT) model,
which is often considered as a prototype of microscopic mechanism for
friction~\cite{mu11}. In the last decades it has been used in theoretical studies
to interpret results from Friction Force Microscopy experiments.  In
PT equations the frictional force $F_C(V)$ acting on the tracer is replaced by
a harmonic force $F_{PT}=-\kappa(X-x)$ linking it to a virtual ``effective'' particle of mass $m$ (whose
position and velocity are denoted by $x,v$ respectively) which
moves on a corrugated surface and is in contact with the environmental
thermostat, see Fig.~\ref{fig:sk}B. In our case the probability $P(X,V,x,v,t)$ obeys Eq.~\eqref{eq:master} with $\Gamma_1=(X,V,x,v)$ and
\begin{subequations} \label{eq: 1}
\begin{align}
H_{}&=\frac{M V^2}{2}+\frac{m v^2}{2}+\kappa\frac{(X-x)^2}{2}+U_0\cos\left(\frac{2\pi x}{L}\right)\\
L_T \cdot &= \left(\frac{\gamma}{m}\right)\frac{\partial (v\cdot)}{\partial v} \cdot + \left(\frac{\gamma k_BT}{m^2}\right) \frac{\partial^2}{\partial v^2} \cdot
\end{align}
\end{subequations}
and $L_{ext}$ is the same appearing in Eq.~\eqref{eq:langevin},
i.e. acting only on $V$ through ``granular'' collisions.  Note that
here for the thermostat we use a Ornstein-Uhlenbeck force of the kind
$-\gamma v+\sqrt{2\gamma T}\xi(t)$, with $\xi(t)$ a Gaussian white
noise.

\begin{figure}
\centering
\includegraphics[width=0.9\columnwidth,clip=true]{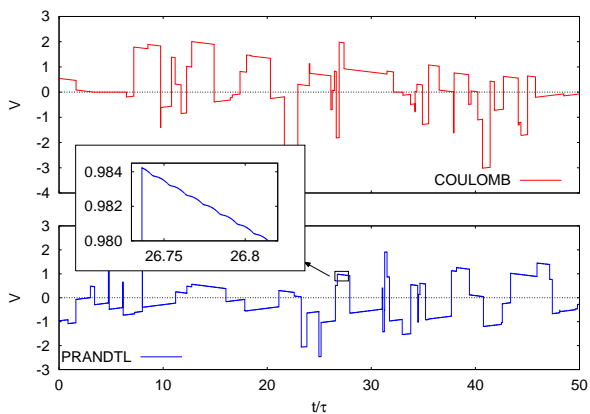}
\caption{Comparison of trajectories from a simulation of the Coulomb
  model (above) and PT model (below). Adimensionalised parameters are: 
  $M_g/M=0.8$ (in both models),  $m/M=10^{-7}$, $\gamma 
  \tau/M=10^{-2}$, $T/T_g=10^{-9}$,
  $U_0/(k_B T_g)=1.6 \cdot 10^{-7}$, $\kappa \lambda^2/(k_B T_g)=10$,
  $L/\lambda=10^{-5}$, which gives $\kappa L^2/(4\pi^2 U_0)=1.6 \cdot
  10^{-4}$. As macroscopic Coulomb force we have used $\Delta
  \tau^2/(M \lambda)=0.05 \sim \langle F_{PT} \rangle$. }
\label{fig:traj}
\end{figure}

The PT force is usually studied in a different context where the first
mass moves at constant velocity, i.e. $X(t)=X(0)+v_0 t$,
reproducing experiments with uniform sliding~\cite{mu11}. In that case, provided
that $k_B T \ll \kappa L^2/2 < 2 \pi^2 U_0$, the stationary state is a
quasi-periodic stick slip motion where, in a range of velocity $v_0
\gg \sqrt{U_0/M}$, the average friction force $\langle F_{PT}
\rangle$ has a negligible (logarithmic or smaller) dependence upon
$v_0$. In our model~\eqref{eq: 1} the velocity of the first mass is
not constant but feels the slowing effect of $F_{PT}$ and, at random
times, is instantaneously changed with probability rates given by
Eq.~\eqref{eq:rates}, i.e. with a typical after-collision value of the
order of $v_g=\sqrt{k_B T_g/M}$. In the model, therefore, it makes
sense to choose $v_g \gg \sqrt{U_0/M}$, that is $k_B T_g \gg U_0$.
When giving the parameters of simulations, we take as unit of mass
$M$, unit of time $\tau$ and unit of length $\lambda=\sqrt{k_B
  T_g/M}\tau$ (that is unit of energy $k_B T_g$).  In
Figure~\ref{fig:traj} we compare the trajectories from a simulation of
the macroscopic model~\eqref{eq:langevin} and a simulation of the
microscopic one~\eqref{eq: 1}. The figure fairly demonstrates that the
trajectories look quite
similar if small details (as in the inset) are ignored.

\begin{figure}
\centering
\includegraphics[width=0.9\columnwidth,clip=true]{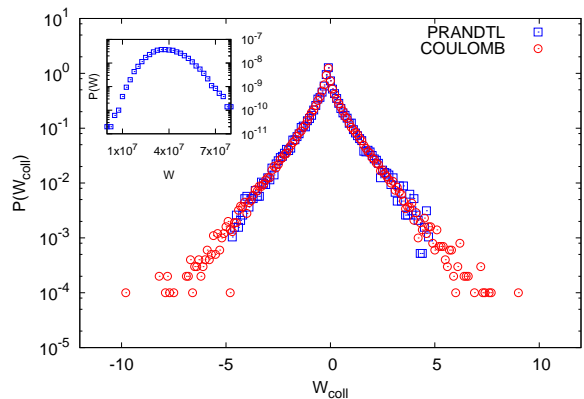}
\caption{Comparison of probability distribution functions of $W_{coll}$
  integrated of a time-window of length $15\tau$, from a simulation of
  the PT model (blue data), and of the Coulomb model (red
  data). Inset: probability distribution function of the action functional $W$ in the PT model. Adimensionalised parameters are $M_g/M=0.8$, $m/M=10^{-7}$, $\gamma \tau/M=10^{-3}$, $T/T_g=2 \cdot 10^{-9}$, $U_0/(k_B
  T_g)= 3.18 \cdot 10^{-7}$, $\kappa \lambda^2/(k_B T_g)=0.1$,
  $L/\lambda=1.4\cdot 10^{-4}$, which gives $\kappa L^2/(4\pi^2 U_0)=1.6 \cdot 10^{-4}$. For the
  macroscopic Coulomb model we used $\Delta
  \tau^2/(M \lambda)=0.0071 \sim \langle F_{PT} \rangle$.  }
\label{fig:pdf}
\end{figure}

Set the model, we now turn our attention to the FEP. The action functional Eq. (\ref{eq:ls}) can be split in two contributions due, respectively, to the collisional and diffusional part of the dynamics $W(t)=W_{coll}+W_{diff}$. Given that in the time interval $[0,t]$ the tracer receives $N_c$ collisions, we have
\begin{equation}
W_{coll}=\sum_{i=1}^{N_c}\ln\frac{k(V_i\to V_i')}{k(-V_i'\to -V_i)}=-\sum_{i=1}^{N_c}\frac{\delta E_i}{k_B T_g}
\end{equation}
with $\delta E_i = \frac{M}{2}(V_i'^2-V_i^2)$ the energy gain in the $i$-th collision.
The term due to diffusion reads
\begin{multline}\label{eq: 5}
W_{diff}=\sum_{i=1}^{N_c}\left(-\int_{t_{i-1}}^{t_i}\,ds\,\mathcal{A}(\Gamma_1(s))+ \right.\\
\left.\int_{t_i}^{t_{i-1}}ds\,\mathcal{A}(\epsilon\Gamma_1 (t_i-s))\right),\nonumber\\
\end{multline}
where 
\begin{equation}
\mathcal{A}(X,V,x,v)=\frac{m}{4\gamma k_BT}\left[m\dot{v}+\gamma v+\frac{\partial H}{\partial x}\right]^2-\frac{\gamma}{2 m}.
\end{equation}
With some algebra one gets
\begin{equation}
W_{diff}=\frac{1}{k_BT}\sum_{i=1}^{N_c}\left[\Delta K_i + \Delta U_i-\int_{t_{i-1}}^{t_{i}}\,ds\,\, \kappa(x-X)V\right],\nonumber\\
\end{equation}
where $\Delta K_i$ and $\Delta U_i$ are the changes of $K=mv^2/2$ and $U=U_0\cos(2\pi x/L) + \kappa(X-x)^2/2$, respectively, in the interval $[t_{i-1},t_i]$. Let us note that, as expected for the unperturbed dynamics, the contribution $W^{i}_{diff}$ to
$W_{diff}$ of a single flight between two collisions at times
$t_{i-1}$ and $t_i$ satisfies detailed balance, i.e.
\begin{equation} \label{eq:diff}
W^i_{diff}=-\frac{\delta H_{i}}{k_B T} \overset{def}{=}\frac{H_{}[\Gamma_1(t_{i-1})]-H_{}[\Gamma_1(t_i)]}{k_B T}.
\end{equation}
A crucial comment is in order concerning the magnitude of the two
contributions to FEP. In the steady state the energies $\delta E_i$
exchanged in the collisions balance, on average, the energies
exchanged with the thermostat, i.e. $\langle \delta E_i \rangle =
-\langle \delta H_{i}\rangle \ge 0$, however $T_g \gg T$ and therefore
$|\langle W_{diff} \rangle|\gg |\langle W_{coll} \rangle|$. In addition the
probability of observing negative values of $W$ (see inset of Fig.~\ref{fig:pdf}) is exceedingly small.
In conclusion, the macroscopic description given by the Coulomb model
completely misses a huge contribution to the FEP due to the coupling
with the environmental thermostat at the temperature $T$. 

The cure of this problem can only come by a correct modelling of the
stochastic part of the thermostat, which is ignored in
$F_C$ of Eq.~\eqref{eq:coulomb}, while it is included in
$F_{PT}$ of Eq.~\eqref{eq: 1}. Model~\eqref{eq: 1} is {\em de facto} a solution of the
problem, as it correctly reproduces both dissipation and fluctuations
due to sliding friction. Its simulation,
however, may require quite an intense computational power if compared with
 Eq.~\eqref{eq:coulomb}. A procedure to complement
Eq.~\eqref{eq:coulomb} with the appropriate stochastic process reproducing the hidden thermostat is indicated
in~\cite{hanggi09}: the explicit form of the noise, however, is not necessary to retrieve the expression for the entropy production. Indeed, if in the absence of external perturbations the dynamics satisfies detailed balance, $W^i_{diff}$ is - by definition - the difference in kinetic energy divided by $T$ (in analogy with Eq. \eqref{eq:diff}).

Leaving aside the problem of entropy production, it is still interesting to observe
the fluctuations of $W_{coll}$ which in principle can be studied in
experiments. These are, apart from the constant $1/(k_B T_g)$ factor,
the fluctuations of the energy flux going from the granular gas into
the tracer. In Figure~\ref{fig:pdf} we show the very good agreement of
the distribution of these fluctuations in simulated steady states of the two models.  Obtaining general
relations for the fluctuations of $W_{coll}$ remains an open problem:
a starting point is offered by the known relations for the joint
probability distribution of currents~\cite{gaspard07}.

\section{Conclusion}

In this Letter we have considered the effect of coarse-graining which
is operated when dealing with out-of-equilibrium macroscopic systems,
where the interaction with environmental thermostats is replaced by
effective (often phenomenological) dissipative forces. We have shown
how the effect of coarse-graining drastically changes the properties
of the model under time-reversal if fluctuations are not properly
described. In the case of Coulomb friction, the friction force
originates from a configurational potential but is modelled as a
velocity-dependent force, changing its time-reversal parity. Moreover,
if the Coulomb law is introduced without its conjugate fluctuations,
the dominant part of the FEP is lost.

The general idea, as always, is that coarse-graining is a loss of
variables, or information. Such information is relevant or not,
depending on the question one
considers~\cite{pigolo,Crisanti2012}. Microscopic variables are not
really relevant for many observables: for instance, in the above
example, the correct fluctuations of $W_{coll}$ are perfectly
recovered even in the coarse-grained model. Other properties, which
require a finer knowledge of the system, are lost.

A similar situation has been encountered,
in the past, with inelastic collisions: the collisional dissipation is
modelled as an instantaneous loss of energy without thermal
fluctuations~\cite{PVBTW05}. The real FEP, therefore,
cannot be accounted for by the measurement of the dissipated
macroscopic energy in collisions. The dominant channel for entropy
production is the transfer of such energy to the environment, whose
fluctuations are quite difficult to be observed. This situation
sometimes may be counterintuitive, since the energy flux transferred
to the environment is on average the same as that dissipated
macroscopically. However the thermostats involved are totally
different.

\acknowledgments
The authors acknowledge fruitful discussions with A. Sarracino, G. Jona-Lasinio and A. Vulpiani.

\bibliographystyle{eplbib}
\bibliography{mergedbiblio}{}

\end{document}